\documentclass[prl,twocolumn,superscriptaddress,showpacs]{revtex4}
\usepackage{graphicx}
\usepackage{amsmath}
\usepackage{amssymb}
\usepackage{color}

\newcommand{\Tr}{\mathrm{Tr}}

\begin{document}

\title{Explicit Local Integrals of Motion for the Many-Body Localized State}
\author{Louk Rademaker}
\affiliation{Kavli Institute for Theoretical Physics, University of California Santa Barbara, CA 93106, USA}
\author{Miguel Ortu\~{n}o}
\affiliation{Departamento de F\'{i}sica - CIOyN, Universidad de Murcia, Murcia 30.071, Spain}
\date{\today}

\begin{abstract} 
Recently, it has been suggested that the Many-Body Localized phase can be characterized by local integrals of motion. Here we introduce a Hilbert space preserving renormalization scheme that iteratively finds such integrals of motion exactly. Our method is based on the consecutive action of a similarity transformation using displacement operators. We show, as a proof of principle, localization and the delocalization transition in interacting fermion chains with random onsite potentials. Our scheme of consecutive displacement transformations can be used to study Many Body Localization in any dimension, as well as disorder-free Hamiltonians.
\end{abstract}


\maketitle

Since the revival of interest in localization due to disorder\cite{Anderson:1958fz,Basko:2007ta,Basko:2006vz,Nandkishore:2015kt} it has been suggested that the so-called Many-Body Localized (MBL) phase can be characterized by an extensive set of local integrals of motion (LIOM) or $l$-bits, $\tau^z_i$, that commute with each other and the Hamiltonian \cite{2014arXiv1407.8480C,Huse:2014co,2014arXiv1403.7837I,2014arXiv1412.3073K,2014arXiv1406.2175R,Serbyn:2013cl}. Consequently, the Hamiltonian can be written in terms of these LIOMs as
\begin{equation}
	H = \sum_i \xi_i \tau^z_i + \sum_{ij} V_{ij} \tau^z_i \tau^z_j + \ldots.
	\label{LIOM}
\end{equation}
Many properties of the MBL phase, such as its logarithmic entanglement spread or its insulating behavior, can be derived based on this assumption \cite{Serbyn:2013cl}.

The question is, however, what those LIOMs are and how to compute them. Since any sum and product of integrals of motion is itself an integral of motion, the choice of LIOMs is highly arbitrary. Pure mathematically, all projectors onto the (localized) eigenstates are integrals of motion, and out of those one could in principle construct the local integrals of motion. In fact, it is easy to show that all Hamiltonians can be brought into the form dictated by Eqn. (\ref{LIOM}) \cite{LoukInFuture}. As for the MBL phase, Chandran et al.\cite{2014arXiv1407.8480C} use the long-time evolved average of an initially local operator as their LIOMs, whereas Ros et al.\cite{2014arXiv1406.2175R} and Imbrie\cite{2014arXiv1403.7837I} use perturbative methods to construct local integrals of motion.

In this Letter, we construct iteratively a transformation that turns \emph{any} fermionic Hamiltonian into the classical form of  Eqn. (\ref{LIOM}). This is done by consecutively applying a similarity transformation using a displacement operator $\exp \lambda (X^\dagger - X )$. The elegant properties of this transformation allow for a systematic elimination of off-diagonal interaction terms, order by order in the number of fermionic operators involved. Our renormalization scheme can be used to study Hamiltonians in any dimension, with or without disorder.

Whether a random interacting system is localized or not, depends on how much the integrals of motion $\tau^z_i$ are spread out. As a proof of principle, we apply our method to diagonalize random interacting chains. In both the localized and the delocalized regime we find agreement with Exact Diagonalization. Throughout the phase diagram we find the effective interactions between the integrals of motion, and we can infer the exponential localization of the integrals of motion in the localized regime.

\emph{Definitions - } Here we will consider interacting fermions with random onsite potentials\cite{Basko:2007ta,Basko:2006vz}
\begin{equation}
	H = \sum_\alpha \xi_\alpha c^\dagger_\alpha c_\alpha
	+ \frac{1}{2} \sum_{\alpha \beta \gamma \delta} V_{\alpha \beta \gamma \delta} c^\dagger_{\alpha} c^\dagger_\beta c_\gamma c_\delta,
	\label{FermionModel}
\end{equation}
as it extends the original concept of Anderson localization\cite{Anderson:1958fz} to interacting systems. For this model we will now characterize all the possible terms in the Hamiltonian.

We define a \emph{classical term} in the Hamiltonian as a product of fermionic density operators of the form $n_{i_1} n_{i_2} \ldots n_{i_d}$ where all $i's$ are different. A \emph{quantum term} is the product of fermionic operators that cannot be written as a classical term. It thus contains, next to a possible set of density operators, separate creation and annihilation operators,
\begin{equation}
	X = n_{i_1} n_{i_2} \ldots n_{i_d}
		c^\dagger_{j_1} c_{j_2} c^\dagger_{j_3} c_{j_4} \ldots c_{j_q}. 
	\label{QuantumInt}
\end{equation}
Again, all the $i$'s and all the $j$'s are different from each other. The \emph{order} $\mathcal{O}(X)$ of a term is defined as the total number of creation and annihilation operators it contains. This equals $\mathcal{O}(X)~=~2d+q$ where $d$ is the number of density operators and $q$ the number of separate creation and annihilation operators. We require the order to be even, and the interaction to contain the same number of creation and annihilation operators, so that all interactions preserve the total fermion number\footnote{This constraint restricts the application of our method. Examples of models that do not conserve the total fermion number are superconductivity theories, Majorana fermion models or models with quasi-electrons and holes that can annihilate each other.}. We call the notation of Eqn. (\ref{QuantumInt}) \emph{normal ordered}: all density interactions are grouped together, and the remaining is a product of alternating creation and annihilation operators which all act on different sites.

The order of a sum of terms is defined as the minimum of the orders of the individual terms,
\begin{equation}
	\mathcal{O} \left( \sum_i X_i \right) 
		=  \min_i \left( \mathcal{O}(X_i) \right).
\end{equation}

When multiplying two terms $X$ and $Y$, the product $XY$ can contain interaction terms of order lower than $\mathcal{O}(X) + \mathcal{O}(Y)$. This happens when $X$ contains the annihilation operator on site $\alpha$ and $Y$ contains the creation operator on the same site, we call this an \emph{overlap}. This gives $c_\alpha c^\dagger_\alpha = 1 - n_\alpha$, which generates a term of an order two lower. Since this can happen for any pair of creation and annihilation operators in $X$ and $Y$, there exists a case of \emph{maximal overlap} with overlap on $\frac{1}{2} \min(\mathcal{O}(X), \mathcal{O}(Y))$ sites. Because each such overlap generates a term with an order two lower, the product $XY$ has order 
\begin{equation}
	\mathcal{O}(X) + \mathcal{O}(Y) \geq \mathcal{O}(XY) \geq 
		\max \left( \mathcal{O}(X), \mathcal{O}(Y) \right).
\end{equation}

There are three important properties of quantum terms: 1) The product of a quantum term with itself is zero, $ X^\dagger X^\dagger = X X = 0$. 2) The product of a quantum term with its Hermitian conjugate, so $X^\dagger X$ and $X X^\dagger$, is classical. Observe that because $X$ and $X^\dagger$ have maximal overlap, the order remains the same: $\mathcal{O}(X X^\dagger) = \mathcal{O}(X^\dagger X) = \mathcal{O}(X)$. 3) The cubic power is trivial, that is $ X^\dagger X X^\dagger = X^\dagger$ and $X X^\dagger X = X$.

\emph{Displacement Transformation} - Based on the classification of terms we just introduced, we can define a \emph{displacement operator} associated with a quantum term $X$,
\begin{equation}
	\mathcal{D}_X (\lambda) = \exp \left( \lambda (X^\dagger - X ) \right).
\end{equation}
Because of the aforementioned properties of quantum terms, the displacement operator can be written out explicitly as
\begin{equation}
	\mathcal{D}_X (\lambda) =
		1 + \sin \lambda (X^\dagger - X)
			+ (\cos \lambda - 1) (X^\dagger X + X X^\dagger).
	\label{DisplacExplic}
\end{equation}
Note that the Hermitian conjugate of the displacement operator is $D^\dagger_X(\lambda) = D_X(-\lambda)$. The above arguments can easily be extended to spin-$\frac{1}{2}$ Hamiltonians, where classical terms are given by products of $S^z$-operators, and quantum terms are total spin-conserving products of $S^+$, $S^-$ and $S^z$ operators. 

The \emph{displacement transformation} is given by a similarity transformation using the displacement operator. That is, it transforms any term $Y$ as
\begin{equation}
	Y \rightarrow \widetilde{Y} = \mathcal{D}^\dagger_X (\lambda) Y \mathcal{D}_X (\lambda).
\end{equation}
This transformation is similar to a Clifford group rotation, which has transformation operator $\mathcal{D} = e^{\lambda A}$ with $A^2 = 1$, whereas here we have the weaker condition $A^3=-A$.  

For now we use the notation with the tilde to denote the transformed term, we will drop the tilde later as we will perform many consecutive transformations. Under the transformation, there are 'new' terms \emph{generated}, namely $\widetilde{Y} - Y$. Using the explicit formulation of the displacement operator Eqn. (\ref{DisplacExplic}), we see that the 'new' terms are of the form $XY$, $YX$, etc. Carefully counting all the combinations, we see that the order of the new terms is at least the maximum of the orders of $X$ and $Y$,
\begin{equation}
	\mathcal{O}\left( \widetilde{Y} - Y \right)
		\geq \max \left( \mathcal{O}(X), \mathcal{O}(Y) \right).
\end{equation}
As will be shown later, this lower bound on the order of new terms implies the closedness of our systematic transformation procedure.

Without constraining the specific shape of $X$ that we use for the displacement transformation, we can prove that the only way to generate new terms proportional to $X^\dagger + X$ is through terms that have maximal overlap with $X$\cite{OnlineSuppl}. For example, consider an order 4 term $X^\dagger~=~c^\dagger_1 c_2 c^\dagger_3 c_4$ with interaction given by $V$. Let us write down the relevant part of the Hamiltonian as follows
\begin{equation}
	\sum_{i=1}^4 \xi_i n_i + V_{13} n_1 n_3 + V_{24} n_2 n_4 + \frac{1}{2} V (X^\dagger + X).
\end{equation}
The displacement transformation with $\mathcal{D}_X(\lambda)$ leaves the quadratic part untouched, and the prefactor multiplying $(X^\dagger + X)$ becomes
\begin{equation}
	\frac{1}{2} V \cos 2\lambda + \frac{1}{2} \left( \xi_1 + \xi_3 + V_{13} - \xi_2 - \xi_4 - V_{24} \right) \sin 2 \lambda
\end{equation}
so that with $\lambda$ given by
\begin{equation}
	\tan 2 \lambda = - \frac{V}{ \xi_1 + \xi_3 + V_{13} - \xi_2 - \xi_4 - V_{24}}.
	\label{TanLambda}
\end{equation}
the transformed Hamiltonian does no longer have the interaction term $X^\dagger + X$. Similar expressions can be found for transformations involving $X$ of higher order.

The right-hand side of Eqn. (\ref{TanLambda}) equals the 'small' parameter that is used in perturbative studies of MBL\cite{Basko:2007ta,Basko:2006vz,2014arXiv1403.7837I}. Such perturbation theories often run into the problem of \emph{resonances}, where the denominator of the 'small' parameter goes to zero, which means perturbation theory cannot be applied. However, the displacement transformation we present here is well-behaved at a resonance, since then $\lambda = \pi/4$ and the interaction can be still transformed away. Note that each displacement transformation can be viewed as a discrete version of a Wegner transformation\cite{Wegner}.

\emph{Consecutive displacement transformations} - 
Now any fermionic Hamiltonian that respects the total fermion number conservation can be written as
\begin{equation}
	H = \sum_i \xi_i n_i
		+ \sum_{n = 4, 6 \ldots} \sum_j V_{nj} (X_{nj}^\dagger + X_{nj})
\end{equation}
where $\xi_i$ are the onsite energies, $n$ expresses the order of the term $X_{nj}$, $j$ is just an index and $V_{nj}$ are the coupling constants. If all terms $X$ are classical, we have reached our goal: we have a classical Hamiltonian with an infinite set of conserved quantities.

Any quantum term $(X^\dagger + X)$ can be removed from the Hamiltonian by performing a displacement transformation associated with $X$, using the value of $\lambda$ given by Eqn. (\ref{TanLambda}). After done so, we can choose another quantum term and transform that one away - and continue this path of consecutive transformations. 

New terms that are generated are multiplied by either $\sin (\lambda)$, $\cos \lambda$ or products of those. Therefore, generically, new terms have smaller couplings constants, making the process of consecutive transformations alike a renormalization scheme. This generation of new terms takes into account co-tunneling and hence long distance resonances. Note that our scheme preserves the Hilbert space, since each displacement transformation does not decimate sites or bonds nor any other degree of freedom.

In certain cases, however, transforming a term $X$ away can generate terms with an even larger coupling constant. This does not pose a problem: whenever we transform a term $X$ away and it is later regenerated, upon regeneration it will have a smaller coupling constant than before. Additionally, throughout consecutive transformations the distribution of coupling constants will change such that creating larger coupling constants will become more and more unlikely. 
Therefore the magnitude of the strongest coupling constant decreases exponentially with the number of applied transformations\cite{OnlineSuppl}.

We thus remove, term by term, all quantum terms of order 4 in the Hamiltonian. The price we pay is the generation of new terms (both classical and quantum) of order 6 and higher, and new classical terms of order 4. As a result, we obtain a complicated Hamiltonian that is classical in its quadratic and quartic terms. Subsequently, we can do the same tricks for the next order quantum terms, making the Hamiltonian at that level classical as well, and so forth. The procedure cut off at $n$-th order reproduces the exact spectrum for states with $n/2$ particles or less. The computational complexity of diagonalizing a Hamiltonian up to $n$-th order is $\mathcal{O}(N^{3n/2})$, which we confirmed numerically\cite{OnlineSuppl}.

In practice one needs to cut off the procedure at a certain order $n$. This approximation has a clear physical interpretation: we are expressing many-body states systematically in terms of $n$-particle states.

\emph{Numerical implementation} - As a proof of principle, we implemented our method numerically. We consider an open chain of $L$ sites with spinless fermions, with a random onsite energy $\epsilon_i$ on each site chosen uniformly between $-W/2$ and $W/2$, hopping $t=1$ and a nearest neighbor repulsion with $V=1$,
\begin{equation}
	H = \sum_{i=1}^N \epsilon_i n_i 
		+ t \sum_{i=1}^{N-1} (c^\dagger_i c_{i+1} + c^\dagger_{i+1} c_{i} )
		+ V \sum_{i=1}^{N-1} n_i n_{i+1}
	\label{NumericalModel}
\end{equation}
The Hamiltonian is first diagonalized at the quadratic level, after that we continue with displacement transformations at quartic order.

At each iteration we pick the quantum term with the largest coupling constant at a given order, and transform it away. We neglect coupling constants smaller than numerical accuracy, set at $\epsilon = 3 \times 10^{-2}$. The Hamiltonian is thus diagonalized order by order, up to 8th order, yielding 
\begin{equation}
	H = \xi_i \tau^z_i 
		+ V_{ij} \tau^z_i \tau^z_j
		+ V_{ijk} \tau^z_i\tau^z_j\tau^z_k
		+ V_{ijkl}\tau^z_i\tau^z_j\tau^z_k \tau^z_l.
\end{equation}
The results are averaged over 20 disorder realizations for $L=24$ length chains, and 30 realizations at $L=20$.

\begin{figure}
    \includegraphics[width=\columnwidth]{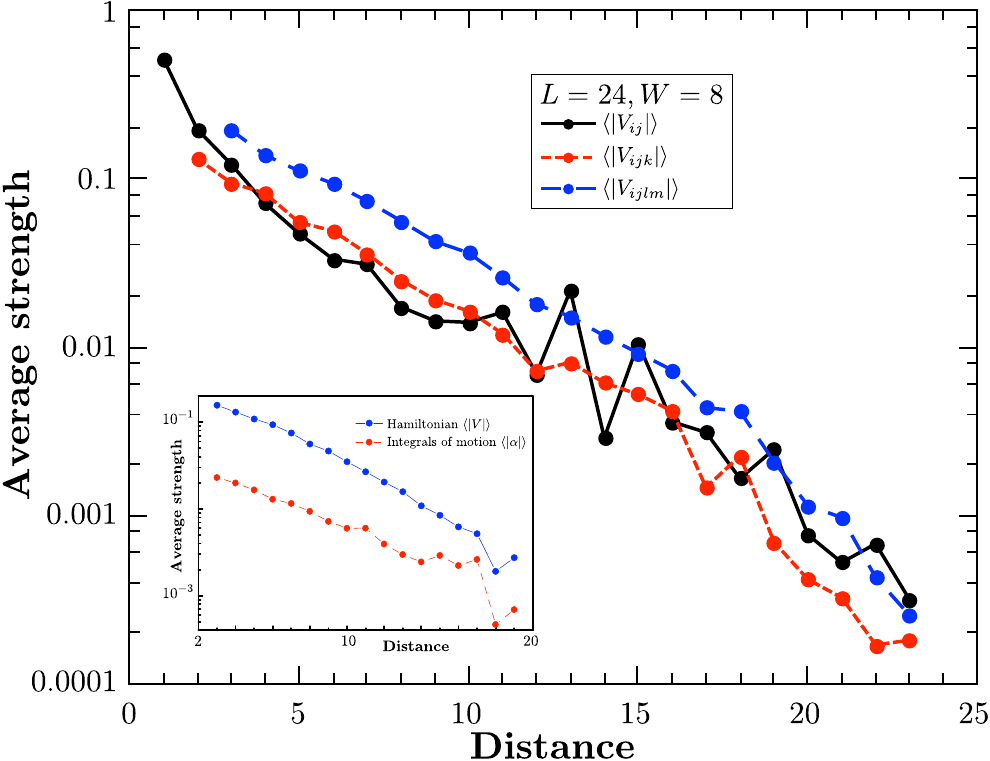}
    \caption{(Color online) The average of the absolute value of the coefficients $\langle |V_{i_1 \ldots i_n}| \rangle$ present in the classical Hamiltonian. Here we show results deep in the many-body localized phase with $W=8$, for a system size $L=24$ averaged over 20 disorder realizations. The different curves represent the two-body $V_{ij}$, three-body $V_{ijk}$ and four-body $V_{ijkl}$ coefficients, respectively. The strength of the coupling constants is almost independent of the order of the interaction. {\bf Inset:} The spread of the local integrals of motion at disorder strength $W=8$ and $L=20$, compared to the coefficients of the Hamiltonian at the same distance. Note the difference between the localization length and the exponential decay of the interactions.}
    \label{Coefficient1}
\end{figure}

\begin{figure}
    \includegraphics[width=\columnwidth]{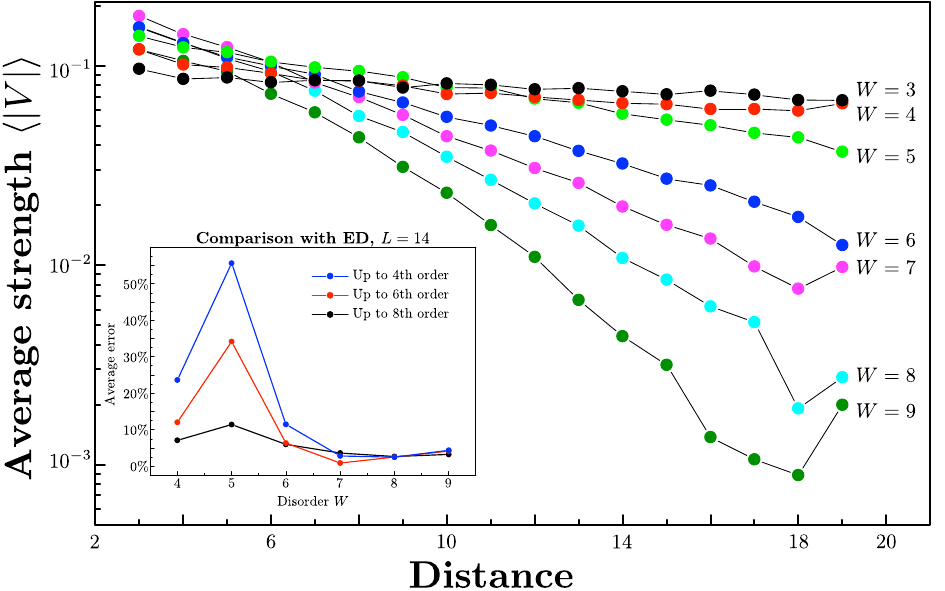}
    \caption{(Color online) The average of the absolute value of the coefficients $\langle |V_{i_1 \ldots i_n}| \rangle$ as a function of distance for various disorder strengths $W=3, \ldots, 9$ for $L=20$. In the localized phase one can see exponential decay, whereas in the delocalized phase at $W < W_c \approx 5$ the coefficients are independent of distance.
    \textbf{Inset:} The relative error in the ground state energy obtained using displacement transformations, compared to exact diagonalization results for $L=14$. The different curves represent the cut-off at a given order $n=4,6,8$. In the localized phase even few-body interactions are sufficient to reproduce the ED results. In the delocalized regime one needs to go up to order 8, whereas around the MBL transition interactions of even higher order need to be included.}
    \label{Coefficient2}
\end{figure}

Deep in the disordered phase, for $W=8$, the coefficients in the Hamiltonian fall off exponentially with the maximum distance of the sites involved. This is quantified by computing the average of the absolute value of the coefficients, $\langle |V_{i_1 \ldots i_n}| \rangle$  as shown in Fig. \ref{Coefficient1}. Note that the strength of the coupling constants appears to be independent of the number of density terms involved. 

One advantage of our method is that we can also compute the classical Hamiltonian in the delocalized phase. In Fig. \ref{Coefficient2} we show the distance-dependence of the Hamiltonian coefficients as a function of disorder, ranging from $W=9$ to $W=3$. If the disorder is weaker than $W_c \approx 5$ the coefficients become independent of distance, which signals delocalization.

To test the accuracy of our method, we compared the ground state energy obtained using our method with Exact Diagonalization (ED) results for $L=14$, see the inset of Fig. \ref{Coefficient2}. With only terms up to 4th order we get a more than 95\% accuracy in the localized phase. However, close to the transition and in the delocalized regime higher order terms are necessary to approach the ED results. For the delocalized regime the 8th order terms seem to be sufficient, yet at the critical point we need even more-body interactions. This suggests our method becomes very expensive close to the localization-delocalization transition.

Having thus established that the practical implementation of our scheme correctly reproduces ED results in the localized regime, we can directly probe the locality of the integrals of motion. The integrals of motion are given by transforming the initial density-operators $n_i$,
\begin{equation}
	\tau^z_i = U^\dagger n_i U = n_i + \sum_{jklm} \alpha^i_{jklm} \; c^\dagger_j c_k c^\dagger_l c_m + \ldots...
\end{equation}
where $U$ is the product of all the displacement transformations. Note that at quadratic order, $\tau^z_i = n_i$, consistent with the fact that the single-particle spectrum is unchanged by the interactions.

The average of the absolute value of the coefficients $\alpha$ as a function of the distance away from the original site $i$ is shown in the inset of Fig. \ref{Coefficient1}. The integrals of motion are indeed exponentially localized, as expected. The localization length for $\tau^z$ is, however, different from the length associated with the decay of the interactions coefficients $V_{ij}$.

These results serve as a proof of principle that the method of displacement transformations can be used to study interacting fermion models. In the Supplementary Information we also applied our method to a different model\cite{OnlineSuppl}. In future work we will apply this method to extract new physical results for various models.

\emph{Outlook -} We introduced a sequence of displacement transformations that allows for the diagonalization of an interacting fermionic Hamiltonian. This method suggests we can bring \emph{any} charge-conserving Hamiltonian into the classical form of Eqn. (\ref{LIOM}), not limited to the many-body localized phase\cite{LoukInFuture} or $d=1$. For example, even the completely nonlocal Fermi liquids\cite{Lai:2015hq} can be analyzed using the classical model\cite{Pines:1999wg}
\begin{equation}
	E = \sum_k \xi_k n_k + \frac{1}{2} \sum_{k k'} f_{k k'} n_k n_{k'} + \ldots
\end{equation}
where the integrals of motion $n_k$ are now local in momentum space and thus delocalized in real space. This begs the question how our method is related to the notion of integrability. We propose that in Eqn. (\ref{LIOM}) quantum integrability depends on the number of parameters $V_{i_1 \ldots i_k}$ that are nonzero, following the definition of Ref. \cite{Caux:2011dx}: if a subexponential or less number of parameters are nonzero and independent, the system is integrable. Note that this case of sparse uncorrelated coupling constants also implies Poissonian level statistics\cite{GuhrRMT}. Our method thus most effectively applies to systems close to integrability. It remains an open interesting question how ergodic systems are represented in the $\tau$-basis of Eqn. (\ref{LIOM}).

A possible fruitful future endeavor would be to recast the iterative transformations in the language of an analytic renormalization scheme, much like the strong disorder renormalization group theory\cite{1994PhRvB..50.3799F,1995PhRvB..51.6411F}, for which a related version has been constructed for MBL systems\cite{Vosk:2013kt,2014arXiv1412.3117V}. A similar method using a Hilbert space preserving RG scheme has been introduced by You~et~al\cite{You2015}. The difference is that they keep interaction terms of all orders, and instead treat the off-diagonal resonance perturbatively to second order. 

Even though the final classical Hamiltonian is of a remarkable simplicity, it does not imply easy solutions since in principle there could be long-range effective interactions between the classical bits $\tau^z$. Furthermore, some complexity of the initial Hamiltonian is transferred to the transformation operator $U$, which is needed to translate any physical operator into the $\tau$-basis. Yet the fact that we can explicitly derive $U$ using the renormalization scheme described in this Letter, introduces a novel quantitative tool for the study of strongly interacting quantum matter.

\emph{Acknowledgments - }
The authors are thankful to Eugeniu Plamadeala, Anushya Chandran, Tarun Grover, Yi-Zhuang You, Cenke Xu, Leon Balents and John Chalker for discussions. L.R. was supported by the Dutch Science Foundation (NWO) through a Rubicon grant. M.O. was supported by Spanish MINECO and FEDER (UE) grant no. FIS2012-38206.



\begin{widetext}

\newpage

\begin{center}
\Large{Online Supplementary Information for \\
\emph{Explicit Local Integrals of Motion for the Many-Body Localized State}}
\end{center}

\section{Maximal overlap with the transformation}

When a displacement transformation is performed with quantum term $X$, only terms with maximal overlap can generate new $X^\dagger+X$ terms. Explicitly, we can enumerate all possible cases:

\begin{enumerate}
	\item A single density term $n_i$ where the site $i$ corresponds to a \emph{creation} operator in $X^\dagger$ transforms as
	\begin{equation}
		n_i \rightarrow n_i + \frac{1}{2} \sin 2 \lambda ( X^\dagger + X)
			- \sin^2 \lambda (X^\dagger X - X X^\dagger).
	\end{equation}
	The product of density terms $n_{i_1} \ldots n_{i_d}$ where \emph{all} the sites $i_1 \ldots i_d$ correspond to creation operators in $X^\dagger$, generate the exact same new terms as the single density term. To prove this, observe that $n_i X^\dagger = X^\dagger$ and $X^\dagger n_i = 0$.
	\item The same results hold, with $-\lambda$ instead of $\lambda$, for a single density term $n_i$ where the site $i$ corresponds to a \emph{annihilation} operator in $X^\dagger$ transforms as
	\begin{equation}
		n_i \rightarrow n_i - \frac{1}{2} \sin 2 \lambda ( X^\dagger + X)
			+ \sin^2 \lambda (X^\dagger X - X X^\dagger).
	\end{equation}
This extends to products of density terms where the sites correspond to only annihilation operators in $X^\dagger$.
	\item The Hermitian interaction $X^\dagger + X$ transforms as
	\begin{equation}
	(X^\dagger + X) \rightarrow
		\cos 2 \lambda (X^\dagger+X)
		- \sin 2 \lambda (X^\dagger X - X X^\dagger).
	\end{equation}
	\item Under the displacement transformation with $X$, new terms of the form $X^\dagger + X$ can \emph{only} be generated by the terms of the shape described in the previous three points. This follows from the fact that in order to generate terms like $X^\dagger + X$, you need to transform a term $Y$ that has order less than or equal to $X$, $\mathcal{O}(Y) \leq \mathcal{O}(X)$, and which has maximal overlap with $X$. This means that \emph{all} creation operators or all annihilation operators in $Y$ should correspond to creation (annihilation) operators present in $X$ or $X^\dagger$. Additionally, there cannot be operators in $Y$ on sites that are not present in $X$, because those would be unaffected by the transition and all new terms would contain operators on these sites. Since $Y$ also contains annihilation operators, they must either live on the same site as the creation operators (density terms), or they live on sites present in $X$. We have thus reduced the possible set of $Y$'s to the three cases presented above.
\end{enumerate}

\section{Diminishing of largest coupling constant}
\label{SecDimLCC}

In certain cases transforming a term $X$ away can generate terms with an even larger coupling constant. We will now show that this does not cause a problem. Explicitly, imagine a Hamiltonian of the form
\begin{equation}
	H^{(0)} = \frac{1}{2} V_X (X^\dagger + X)
	+ \frac{1}{2} V_Y (Y^\dagger + Y)
	+ \frac{1}{2} V_Z (Z^\dagger + Z)
	+ \ldots
\end{equation}
such that $V_X$ is the largest coupling constant, $V_Y,~V_Z~<~V_X$. The terms $Y$ and $Z$ have a maximal overlap with $X$, and maximal overlap with each other. Upon transforming the term $X$ away, the Hamiltonian becomes
\begin{eqnarray}
	H^{(1)} &=&\frac{1}{2} \left( V_Y \cos \lambda_1 + V_Z \sin \lambda_1 \right) (Y^\dagger + Y)
	\nonumber \\ &&
	+ \frac{1}{2} \left( V_Z \cos \lambda_1 - V_Y \sin \lambda_1 \right) (Z^\dagger + Z)
	+ \ldots
\end{eqnarray}
and it is clear that one (not both!) of the new coupling constants can be larger than the original $V_X$. If both coupling constants are smaller, we are contently moving closer to desired convergence. Instead, consider the unfortunate case, where $V_Y \cos \lambda + V_Z \sin \lambda > V_X$. By virtue of our system of consecutive transformations, the next step should be to transform away the term $Y$. Doing so regenerates the original term $X$, however, this time with a smaller coupling constant,
\begin{eqnarray*}
	H^{(2)} &=&
		\frac{1}{2} \left( V_Z \cos \lambda_1 - V_Y \sin \lambda_1 \right) \sin \lambda_2 (X^\dagger + X)
	 \\ &&
	+ \frac{1}{2} \left( V_Z \cos \lambda_1 - V_Y \sin \lambda_1 \right) \cos \lambda_2 (Z^\dagger + Z)
	+ \ldots
\end{eqnarray*}
because $\left( V_Z \cos \lambda_1 - V_Y \sin \lambda_1 \right) \sin \lambda_2 < V_X$ by construction. This implies that the little detour caused by the larger coupling constant has come to an end, and the coupling constant in front of $X$ has been reduced.

Now one can track the magnitude of the coupling strength of each quantum term $X$. Every now and then in the sequence of consecutive displacement transformation the term $X$ has the strongest coupling, and will be transformed away. By the arguments presented above, every next time we transform with $X$ it will have a smaller coupling constant.

\section{Convergence of the method}

As we have shown, the value of each coupling constant reduces through the consecutive application of displacement transformation. We will now estimate the speed at which the method converges, inspired by the works of D. Fisher\cite{1994PhRvB..50.3799F,1995PhRvB..51.6411F}.

For simplicity, let us focus on $n=4$ order terms first, for a system with $L$ sites. The Hamiltonian is, at each step of the renormalization procedure, given by
\begin{equation}
	H^{(i)} = \sum_\alpha \xi_\alpha n_\alpha
		+ \sum_{\alpha \beta} V^{(i)}_{\alpha \beta} n_\alpha n_\beta
		+ \sum_{\alpha \beta \gamma \delta} V^{(i)}_{\alpha \beta \gamma \delta}
			c^\dagger_\alpha c^\dagger_\beta c_\gamma c_\delta
\end{equation}
where the coupling constants $V^{(i)}$ are changing with every transformation. The total number of independent quantum terms are
\begin{equation}
	N_{Q} = \frac{1}{8} (L+1) L (L-1)(L-2).
\end{equation}
Additionally, there are $\frac{1}{2} L (L-1)$ classical terms at order $n=4$, and $L$ classical terms at order $n=2$ that won't change under renormalization.

In the spirit of strong disorder renormalization group theory (SDRG), we define a probability distribution $P_i(X)$ for the absolute value of the quantum interaction coupling constants $X = |V^{(i)}_{\alpha \beta \gamma \delta}|$. Since we are explicitly looking at finite size systems, we must be careful about the interpretation of a probability distribution. A finite size system can be viewed as a specific realization with $N_Q$ terms taken from the distribution $P_i(X)$. Given a continuous $P_i(X)$, the expectation value for largest coupling constant becomes
\begin{equation}
	\Gamma_i = N_{Q} \int_0^\infty dX  \; X \; P_i (X) \left( \int_0^X dY P_i(Y) \right)^{N_{Q} - 1}.
	\label{LargestCouplC}
\end{equation}
For example, if $P_i(X)$ is a uniform distribution between $[0,1]$, the expectation value for $\Gamma_i = 1 - \frac{1}{N_Q} + \mathcal{O}(L^{-5})$. Similarly, the expectation value for the second-to-largest coupling constant is 
\begin{equation}
	\Gamma'_i = N_{Q} (N_Q-1) \int_0^\infty dX  \; X \; P_i (X) \left( \int_0^X dY P_i(Y) \right)^{N_{Q} - 2} \left( \int_X^\infty dY P_i(Y) \right).
\end{equation}
The uniform distribution yields $\Gamma_i' = 1 - \frac{2}{N_Q} + \mathcal{O}(L^{-5})$, which is the expected result.

Each displacement transformation changes the distribution $P_i(X)$. For a finite size system, we must pick the largest realized coupling constant $|V^{(i)}_{\alpha \beta \gamma \delta}| =\Gamma_i$, and apply the transformation. This coupling constant becomes zero. The new distribution, by virtue of the fact that $\Gamma_i$ was the largest realized coupling constant, should have zero weight for $X > \Gamma_i$. We obtain the distribution of the remaining couplings,
\begin{equation}
	\widetilde{P}_i(X)  = P_i(X) - \Theta( X - \Gamma_i) P_i(X) 
			+ \left( \int_{\Gamma_i}^\infty P_i(Y) dY \right) \delta(X).
		\label{Step1}
\end{equation}
The next step is to see how the given displacement transformation acts on the remaining terms. Only terms with maximal overlap will be able to generate new terms at the same order. Given any interaction there are $N_{overlap} = (L-4)(L-5)/2$ terms that have maximal overlap. As discussed in Sec. \ref{SecDimLCC}, the overlapping terms always come in pairs, say $Y$ and $Z$. The old interaction strengths $V_Y$ and $V_Z$ are transformed into $V_Y \cos \lambda + V_Z \sin \lambda$ and $V_Z \cos \lambda - V_Y \sin \lambda$. Consequently, a fraction $N_{overlap}/N_Q$ is changed by the transformation, and the new distribution function of couplings becomes
\begin{equation}
	P_{i+1}(X) = \widetilde{P}_i(X)  \left( 1 - \frac{N_{overlap}}{N_Q} \right)
			+ \frac{N_{overlap}}{N_Q} 
			\frac{\int_0^\infty dZ \; \widetilde{P}_i(\frac{X-Z}{\cos \lambda_i})   
				\widetilde{P}_i(\frac{Z}{\sin \lambda_i}) }{\cos \lambda_i \sin \lambda_i}
	\label{Step2}
\end{equation}
where $\lambda_i$ is the parameter associated with the displacement transformation.

The remaining $N_{Q} - N_{overlap} - 1$ terms are unchanged, hence Eqn. (\ref{Step1}) and Eqn. (\ref{Step2}) represent the change in the distribution function due to one displacement transformation.

The method of consecutive displacement transformations converges if the expectation value of the largest coupling constant under the new distribution $P_{i+1}(X)$ reduces significantly. We will now show that if initially only a sparse set of interactions have 'large' coupling constants, the largest coupling constant reduces exponentially. If, on the other hand, the system has initially a large set of large coupling constants, the distribution $P_i(X)$ will naturally become more and more sparse close to its upper bound. Once a sparse distribution is reached, the first argument again applies and a exponential decay sets in.

\subsection{Initially sparse set of interactions}
Consider a system where, initially, only $\mathcal{O}(L^2)$ of the quantum terms have a nonzero coupling constant. A typical initial distribution would be
\begin{equation}
	P_0 (X) = \left( 1 - \frac{1}{L^2} \right) \delta(X) + \frac{1}{L^2 \Lambda_0} \Theta (\Lambda_0 - X).
\end{equation}
In the large $L$ limit, the expectation value for the largest coupling constant is
\begin{equation}
	\Gamma_0 = \Lambda_0 \left( 1 - \frac{1}{L^2} \right).
\end{equation}
Let us now apply Eqns. (\ref{Step1}) and (\ref{Step2}) to this distribution. The first step amounts to removing the largest coupling constant. We verify explicitly that we moved one coupling constant ($\mathcal{O}(L^{-4})$ of the total number of terms) to zero strength,
\begin{equation}
	\widetilde{P}_0(X) = \left( 1 - \frac{1}{L^2} + \frac{1}{L^4} \right) \delta(X) + \frac{1}{L^2 \Gamma_0} \left( 1- \frac{1}{L^2} \right) \Theta (\Gamma_0 - X).
\end{equation}
The next step requires knowledge of the transformation parameter $\lambda_0$, which enters into the convolution of $\widetilde{P}_0$ with itself, see Eqn. (\ref{Step2}). The self-convolution of $\widetilde{P}_0(X)$ gives for $X>0$ explicitly
\begin{eqnarray}
	&&\frac{1}{\cos \lambda_0 \sin \lambda_0} \int_0^\infty dZ \; 
	\widetilde{P}_0 \left(\frac{X-Z}{\cos \lambda_0} \right)   \widetilde{P}_0
	\left(\frac{Z}{\sin \lambda_0} \right)
	 \nonumber \\ &&
		=\frac{A_0(1-A_0)}{\Gamma_0} \left[ \frac{\Theta( \Gamma_0 \sin \lambda_0 - X )}{\sin \lambda_0} + \frac{\Theta( \Gamma_0 \cos \lambda_0 - X )}{\cos \lambda_0} \right] 
		\nonumber \\ &&
	+ \frac{A_0^2}{\Gamma_0^2 \sin \lambda_0 \cos \lambda_0} \left\{ 
	\begin{array}{ll}
	X, & 0 < X < \Gamma_0 \sin \lambda_0; \\
	\Gamma_0 \sin \lambda_0, & \Gamma_0 \sin \lambda_0 < X < \Gamma_0 \cos \lambda_0 ;\\	
	\Gamma_0 (\sin \lambda_0 + \cos \lambda_0 ) - X, & \Gamma_0 \cos \lambda_0 < X < \Gamma_0 (\cos \lambda_0 + \sin \lambda_0); \\
	0 & \mathrm{else}
	\end{array}
	\right.
	\label{Convo1}
\end{eqnarray}
with $A_0 = \frac{1}{L^2} - \frac{1}{L^4}$. The terms have a simple physical explanation. The first term with $\sin \lambda_0$ is obtained because new terms are generated that had originally $V=0$, but due to the transformation have become nonzero. The next term consists of terms that were originally present, but due to the transformation are reduced by a factor $\cos \lambda_0 \leq 1/\sqrt{2}$. In both cases, it affects only a relative $\mathcal{O}(L^{-2})$ number of terms. The final term on the last line consists of the pairs as discussed in Sec. \ref{SecDimLCC}, where both $V_Y$ and $V_Z$ were nonzero prior to the transformation.

This last part is dangerous, since it generates a weight of the distribution for $X$ larger than $\Gamma_0$. However, even in the worst case scenario of a many-body resonance, which yields $\lambda_0 = \pi/4$, the integrated probability to find a new coupling constant larger than $\Gamma_0$ is
\begin{equation}
	\frac{A_0^2 N_{overlap}}{ N_Q} (\sqrt{2} - 1)^2
	\approx \frac{0.68}{L^6} + \ldots
\end{equation}
We are saved by the sparseness of the initial set! In the new distribution $P_1(X)$ the expectation value for the largest coupling constant is, in the large $L$ limit,
\begin{equation}
	\Gamma_1 \approx \Gamma_0 \left( 1 - \frac{1}{L^2} \right).
\end{equation}
Since the top of the distribution remains $\mathcal{O}(L^{-2})$ with our method, this estimate can be generalized to every step. Consequently, the coupling constant after $i$ steps is given by
\begin{equation}
	\Gamma_i \sim \left(1 - \frac{1}{L^2} \right)^i \sim \exp \left( - i / L^2 \right).
	\label{GammaSmaller}
\end{equation}
We have shown that for initially sparse interactions, the largest coupling constant will be reduced exponentially. This is consistent with, for example, the inset of Fig. \ref{Localization36} of this Supplementary Information.

Note that Eqn. (\ref{GammaSmaller}) implies that the number of transformations required scales in a polynomial fashion with system size. The overall computational complexity of diagonalizing a Hamiltonian using displacement transformations, at a given order, is therefore also polynomial. 

\subsection{Initially dense set of interactions}

The above arguments do not apply when our initial distribution has a dense set of large interactions. A typical example is the uniform distribution,
\begin{equation}
	P_0(X) = \frac{1}{\Lambda_0} \Theta( \Lambda_0 - X)
\end{equation}
The expectation value for the largest coupling constant is now, $\Gamma_0 = \Lambda_0 \left( 1 - \frac{1}{L^4} \right).$ From the uniform distribution it is clear that there will be a macroscopic number of quantum terms that will, through the transformation, get a coupling constant {\em larger} than $\Gamma_0$. As Eqn. (\ref{Convo1}) shows, the new distribution function has a cut-off at $\Lambda_1 = \Gamma_0 ( \cos \lambda_0 + \sin \lambda_0)$ and it goes to zero in a linear fashion, $P_1(X) \sim (\Lambda_1 - X)$.

The self-convolution of $P_1$ will yield a distribution that will go to zero as $P_2(X) \sim (\Lambda_2 - X)^3$. In general, if $P_i(X)$ goes to zero as $(\Lambda_i - X)^\beta$, the next distribution has a tail of the form $P_{i+1}(X) \sim (\Lambda_{i+1} - X)^{2\beta+1}$. The central limit theorem tells us that quickly the distribution will approach a normal distribution.

A normal distribution is a prime example of a distribution with a sparse tail. Once the distribution $P_i(X)$ has reached its Gaussian shape, we can revert to the arguments of the previous subsection to claim an exponential decay of the largest coupling constant.

Notice that this argument is very similar to Fishers a posteriori justification for SDRG by introducing the infinite disorder fixed point\cite{1994PhRvB..50.3799F,1995PhRvB..51.6411F}. Here the displacement transformation method seems to fail when there are many 'large' interactions, however, the consecutive transformations make the tail of the coupling constant distribution so small that we reach a regime of exponentially sparse 'large' coupling constants.

\subsection{Computational complexity}

We have just shown that for order $n=4$ the largest coupling constant falls off exponentially with prefactor $L^2$. For higher order interactions we observe that the relative density of terms with maximal overlap, $N_{overlap}/N_Q$, scales as $L^{-n/2}$. Consequently, to diagonalize a Hamiltonian up to order $n$ given a fixed numerical precision, one needs to perform $L^{n/2}$ displacement transformations. Because there are $~L^n$ quantum terms up to order $n$, the total computational complexity scales as
\begin{equation}
	\mathrm{CPU \; time} \sim L^{3n/2}.
\end{equation}

We computed the actual CPU time needed to diagonalize the model introduced in the main text, up to order $n=4$, as a function of different system sizes. The result is shown in Fig. \ref{CPUTime}, which indeed satisfies a polynomial scaling in system size with the correct power $L^6$.

\begin{figure}
    \includegraphics[width=0.4\columnwidth]{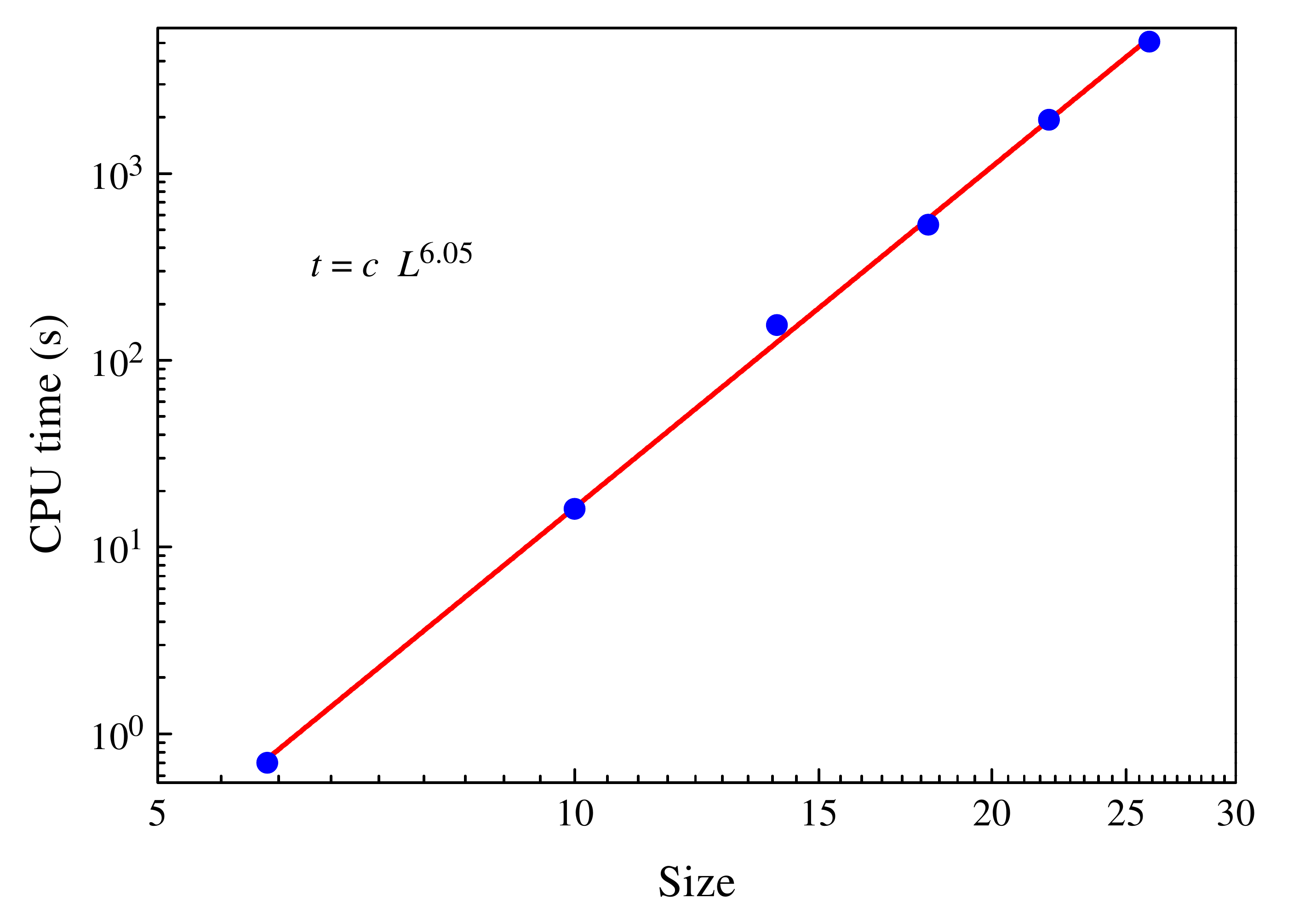}
    \caption{The computational time required to diagonalize the model of the main text up to order $n=4$. We find polynomial dependence on the system size, with $\mathrm{CPU \; time} \sim L^{3n/2}$. }
    \label{CPUTime}
\end{figure}

\section{Simple test-model}

In the main manuscript we study the Anderson insulator with nearest neighbor interactions. Here we briefly discuss another, simpler model that we tested our method for. Consider a periodic chain of $N$ sites with spinless fermions, with a random chemical potential $\xi_i$ on each site chosen uniformly between $-W$ and $W$. The interactions couple four neighboring sites, with uniform strength $V$. The Hamiltonian is
\begin{equation}
	H = \sum_{i} \xi_i n_i + \frac{V}{2} \sum_i \left( c^\dagger_i c_{i+1} c^\dagger_{i+2} c_{i+3} + \mathrm{h.c.} \right).
	\label{NumericalModel}
\end{equation}
One expects a localization-delocalization transition as a function of disorder strength $W/V$. 

We used a different code than used in the main manuscript. At each step we pick the quantum term with the largest coupling constant, and transforms it away. We neglect coupling constants smaller than numerical accuracy, set at $\epsilon = 10^{-12}$ in units where $V=1$. To speed up the computation, each step we throw away all terms of order 6 and higher. With this procedure, we indeed find that the magnitude of the largest coupling decreases rapidly, as shown in the inset of Fig. \ref{Localization36}.

After order $N^2$ iterations, we have realized the classical Hamiltonian $\widetilde{H} = \sum_i \xi \widetilde{n}_i + V_{ij} \widetilde{n}_i \widetilde{n}_j$. Within the model Eqn. (\ref{NumericalModel}), only next-nearest neighbor interactions are generated, $V_{ij} \sim \delta_{|i-j| = 2}$. In this model, the structure of $V_{ij}$ is therefore not very enlightening to study the MBL phase.

A better measure of the localization is to directly probe the locality of the new integrals of motion. To do so, we start out with a density operator $n_i$ on a site, and transform it using the same transformations that diagonalized the Hamiltonian. The results will be of the form, up to quartic order,
\begin{equation}
	\tau^z_i = U^\dagger n_i U = n_i + \sum_{jklm} \alpha^i_{jklm} \; c^\dagger_j c_k c^\dagger_l c_m
\end{equation}
where $U$ is the product of all the displacement transformations. Note that at quadratic order, $\tau^z_i = n_i$, consistent with the fact that the single-particle spectrum is unchanged by the interactions.

Now there are various methods of determining whether a $\tau^z_i$ is quasi-local. Ref. [\onlinecite{2014arXiv1407.8480C}] suggested to use the infinite-temperature overlap between IOMs at different sites,
\begin{equation}
	M_{ij} = 4  \Tr( \tau_i^z \tau_j^z ) - 1
\end{equation}
which is shown in Fig. \ref{Localization12} for a $N=12$ chain for two different disorder strengths. One sees a clear indication of localization in the case of strong disorder $W$, and delocalization in the case of weak disorder.

Another method, which is less time-consuming as it does not involve any trace, directly sums for each distance $d$ the absolute value of the prefactors $|\alpha|$ of terms that act on sites at distance $d$ from each other. We computed this IOM spread on a $N=36$ length chain, which is larger than state-of-the-art Exact Diagonalization studies can reach. The results of is shown in Fig. \ref{Localization36} for two values of the disorder strength. Indeed, for strong disorder case we find localization whereas for weak disorder the IOMs have weight throughout the full length of the chain.

\begin{figure}
    \includegraphics[width=0.4\columnwidth]{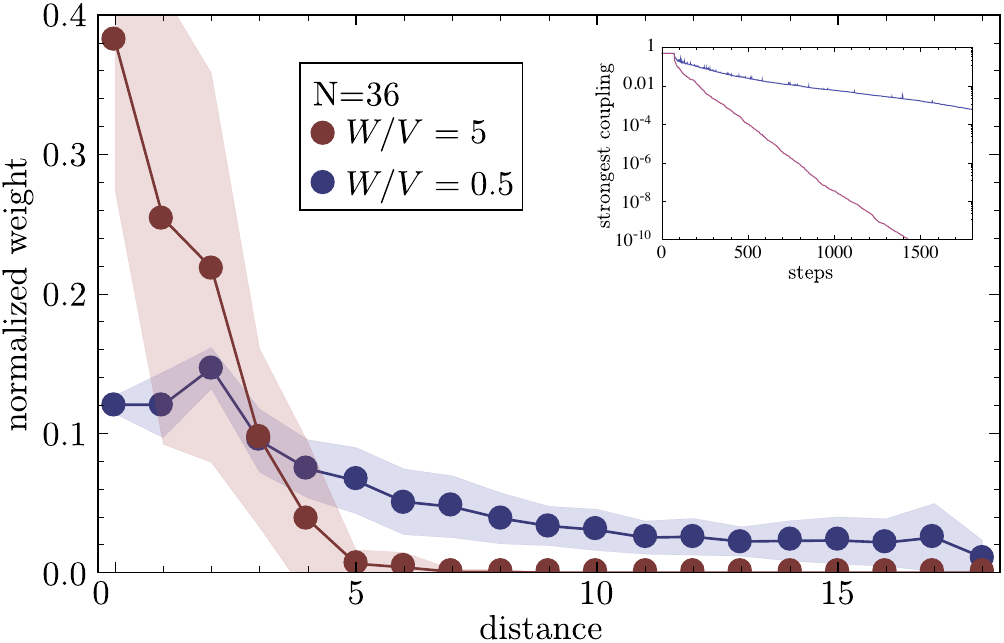}
    \caption{The normalized spread of the local integrals of motion for a $N=36$ chain with $W/V=5$ and $W/V=0.5$. The shaded area represents the standard deviation when averaging over all integrals of motion. The curve is normalized, so that the area under the curve equals one. \textbf{Inset:} The magnitude of the strongest coupling as a function of the number of transformations. It decreases exponentially, though much slower in the delocalized phase than in the localized phase.}
    \label{Localization36}
\end{figure}

\begin{figure}
    \includegraphics[width=0.4\columnwidth]{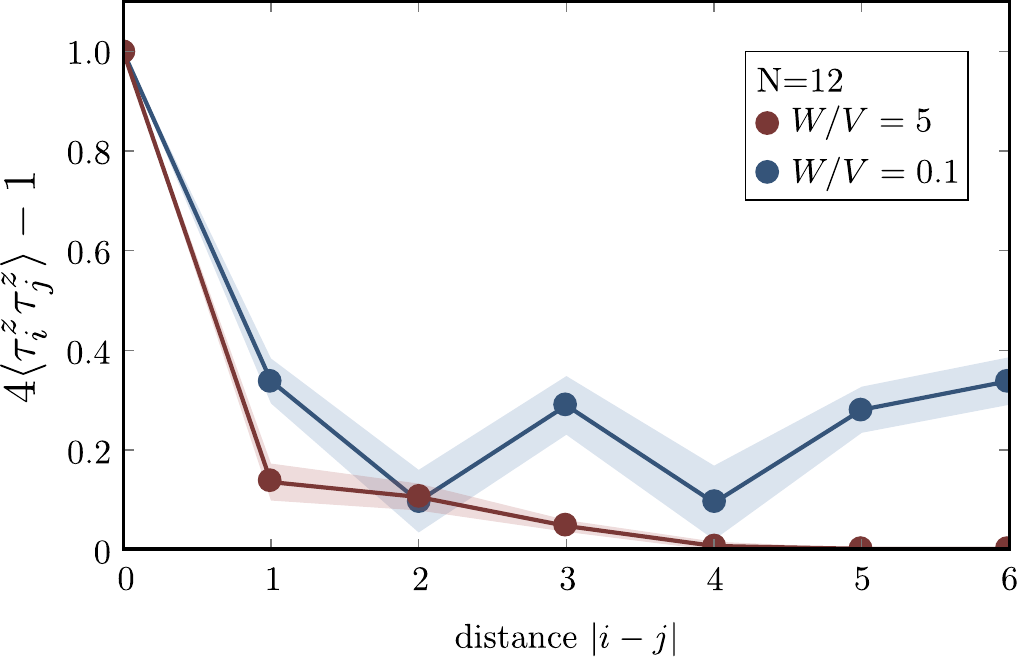}
    \caption{The correlation function $M_{ij} = 4 \Tr ( \tau^z_i \tau^z_j ) - 1$ as a function of distance, averaged over disorder realizations. This function expresses the localization of the local integrals of motion for a $N=12$ chain with $W/V=5$ and $W/V=0.1$. The shaded area represents the standard deviation.}
    \label{Localization12}
\end{figure}

\end{widetext}

\end{document}